\documentclass[12pt]{article}

\usepackage{amsmath,amsthm,amsfonts,amssymb,bbm,bm}
\usepackage{graphicx,wrapfig,float,url,natbib,booktabs,caption,subcaption,comment,multirow,color,tikz,setspace,wrapfig,hyperref,rotating,enumitem}
\usepackage{algorithm,algpseudocode}

\newtheorem{theorem}{Theorem}

\newcommand{\Hcal}{\mathcal{H}}
\newcommand{\Ucal}{\mathcal{U}}
\newcommand{\Ebb}{\mathbb{E}}
\newcommand{\Pbb}{\mathbb{P}}
\newcommand{\Rbb}{\mathbb{R}}

\newcommand{\var}{\mathrm{var}}
\newcommand{\cov}{\mathrm{cov}}
\newcommand{\tr}{\mathrm{tr}}
\newcommand\diag{\mbox{diag}}
\newcommand{\trans}{^{\mbox{\tiny{\sf  T}}}}

\newcommand{\blind}{1}

\begin{document}

\if1\blind
{
\begin{center}
{\Large{\bf Estimation and Inference for High-dimensional \\ }}
\medskip
{\Large{\bf Multi-response Growth Curve Model}} \\
\vspace{0.25in}
		
{\large{\sc Xin Zhou$^1$, Yin Xia$^2$\footnote{Address for correspondence: Yin Xia, Department of Statistics and Data Science, School of Management, Fudan University, 670 Guoshun Road, Siyuan Building, Room 502. Shanghai 200433, P.R. China. E-mail: xiayin@fudan.edu.cn.},  and Lexin Li$^1$}}\\
\vspace{0.15in}
{\it $^1$University of California at Berkeley, and $^2$Fudan University}
\vspace{0.2in}
\end{center}
} \fi

\if0\blind
{
\begin{center}
{\Large{\bf Estimation and Inference for High-dimensional \\ }}
\medskip
{\Large{\bf Multi-response Growth Curve Model}} \\
\vspace{1.025in}
\end{center}
} \fi

\bigskip
\begin{abstract}
A growth curve model (GCM) aims to characterize how an outcome variable evolves, develops and grows as a function of time, along with other predictors. It provides a particularly useful framework to model growth trend in longitudinal data. However, the estimation and inference of GCM with a large number of response variables faces numerous challenges, and remains underdeveloped. In this article, we study the high-dimensional multivariate-response linear GCM, and develop the corresponding estimation and inference procedures. Our proposal is far from a straightforward extension, and involves several innovative components. Specifically, we introduce a Kronecker product structure, which allows us to effectively decompose a very large covariance matrix, and to pool the correlated samples to improve the estimation accuracy. We devise a highly non-trivial multi-step estimation approach to estimate the individual covariance components separately and effectively. We also develop rigorous statistical inference procedures to test both the global effects and the individual effects, and establish the size and power properties, as well as the proper false discovery control. We demonstrate the effectiveness of the new method through both intensive simulations, and the analysis of a longitudinal neuroimaging data for Alzheimer's disease. 
\end{abstract}
\bigskip

\vfill

\newpage
\baselineskip=24pt

\section{Introduction}
\label{sec:introduction}

A growth curve model (GCM) describes how an outcome variable evolves, develops and grows as a function of time along with other related covariates, and is particularly useful for modeling the growth trends in longitudinal data analyses. In a GCM, each individual subject is assumed to have her own unique trajectory of change over time, which  represents how the outcome evolves for each subject as time progresses. Such individual trajectories are modeled as functional curves, typically in the form of linear functions. The model involves both fixed-effects that represent population-level relation between the predictors and outcome, as well as random-effects that account for individual variability in the growth trajectories. These random-effects allow for individual differences in both the starting point (intercept) and the rate of change (slope) over time, and capture the deviations of each individual's trajectory from the average trajectory. GCM has been commonly used in a wide range of scientific applications, for instance, biology, psychology, social science, neuroscience, among others \citep{Wei2006, curran2012multivariate, newhill2012growth, nocentini2013level, ordaz2013longitudinal}. 

Our motivation arises from longitudinal studies of Alzheimer's disease (AD). AD is an irreversible neurodegenerative disorder and the leading form of dementia in elderly subjects. It is characterized by progressive impairment of cognitive and memory functions, then loss of independent living, and ultimately death. Its prevalence is rapidly growing as the worldwide population is aging, and it becomes imperative to understand, diagnose, and treat this disorder \citep{Jagust2018-review}. In recent years, a number of longitudinal AD studies are emerging to track and better understand the progression of AD. One example is \citet{Marcus2010-OASIS2}, who collected T1-weighted magnetic resonance imaging (MRI) scans of 150 subjects aged between 60 to 96 years old. Each subject was scanned on two or more visits, separated by at least one year, for a total of 373 image scans. Each MRI scan was preprocessed, mapped to a common brain atlas, and summarized as a vector of region-wise brain volume measurements. Among those subjects, 64 were characterized as demented at the time of their initial visits and remained so for subsequent scans, and 72 were normal developing controls. A scientific question of central interest is to track the change of brain volumes of different brain regions across time, and understand the difference of developmental trajectories between the AD patients and normal controls. Another example is \citet{Zhong2020}, who collected blood plasma samples of 35 subjects over 70 years old and followed them up to 15 years. Each subject were sampled three times or more, for a total of 164 plasma samples. For each sample, the extracellular RNA (exRNA) sequence measurements were obtained. Among those subjects, 15 were confirmed AD patients by pathological analysis of their post mortem brains, and 9 were normal controls. A central scientific question is to track the change of exRNA expression levels of a set of AD-related genes over time, and differentiate their developmental trajectories between the two groups of subjects. Such questions are pivotal for our understanding of AD development, and have important diagnostic and therapeutic implications. For both examples, GCM provides a natural framework to address the scientific questions of interest, which translate to the estimation and inference of the corresponding fixed-effects parameters in the model. Meanwhile, it is crucial to take into account the spatial correlations between different brain regions or genes, the temporal correlations across different time points, as well as the individual subject variability. 

In this article, we study a high-dimensional multivariate-response linear GCM, and develop the corresponding estimation and inference procedures. Our proposal directly addresses the questions in our motivating examples, where the brain regions or genes are modeled as the multivariate response variables. Although GCM is a classical model, the estimation and inference of a high-dimensional multi-response GCM faces numerous serious challenges, and remains underdeveloped. First, the dimension of the response variables can be large, resulting in a very large covariance matrix. Meanwhile, the number of subjects and the number of time points may be limited. To obtain a good covariance estimator, we adopt a Kronecker product structure for the covariance, which allows us to pool the correlated samples effectively to improve the estimation accuracy. Second, the covariance structure is particularly complex. Under the Kronecker product assumption, the covariance involves three key components, including a spatial covariance that accounts for the correlations among different regions or genes, a temporal covariance that accounts for the correlations across multiple time points, and a covariance matrix that reflects the random departure of individual subjects from the population-level intercept and slope. We devise a highly non-trivial multi-step approach to estimate the three covariance components separately and effectively. Last but not least, we develop rigorous statistical inference procedures to test both the global effects and the individual effects, and we establish the size and power properties, as well as the proper false discovery control. We first recognize that our covariance estimator is not the usual sample covariance, because the means of the response variables in our setting are different for different subjects, regions and time points. Therefore, the asymptotics of the classical sample covariance estimator based on independent and identically distributed (i.i.d.) observations are not directly applicable. To overcome this issue, we introduce some level of sparsity, and show that our estimator performs asymptotically the same as the sample covariance matrix obtained by the i.i.d.\ centralized error terms. We then employ an advanced version of Hanson-Wright inequality \citep{chen2021testing}, and derive the proper convergence rates of our covariance estimators, which in turn ensure the desired theoretical guarantees for our proposed tests. 

Our proposal is built upon but is also clearly different from several lines of relevant research. The first line is classical GCM modeling, which often adopts the approach of linear multilevel analysis. There have been numerous proposals for GCM estimation \citep{kackar1981unbiasedness, goldstein1986multilevel, bryk1992hierarchical}, and inference \citep{giesbrecht1985two, kenward1997small, carpenter1999non, li2015testing}. However, most of the classical GCM solutions focus on a univariate response, or a low and fixed dimensional response scenario, and none tackles the global testing and multiple testing problems simultaneously. By contrast, we target both types of testing problems in a high-dimensional response scenario. The second line of related research is inference for high-dimensional linear mixed-effects model, since GCM can essentially be rewritten as a mixed-effects model. There have been a number of recent proposals toward this end \citep{bradic2020fixed, law2022inference, li2022inference}. However, they mostly focus on testing a single fixed-effect coefficient, and none considers the global and multiple testing problems either. Moreover, those approaches usually obtain an estimator of the fixed-effects through a proxy of the covariance matrix, which may result in efficiency loss. In comparison, we borrow the region and time information to obtain an appropriate estimator for the true covariance matrix, which improves the power of the subsequent inferences. The third line of related research is high-dimensional inference, including the global and multiple testing procedures \citep{cai2013two,cai2014two,liu2013ggm,xia2018multiple}, and the testing procedures involving a Kronecker product structure \citep{xia2017hypothesis, xia2019matrix, chen2019graph, chen2021testing}. We also adopt the Kronecker product structure to facilitate our analysis. But we target a completely different problem from the existing solutions. As a result, we develop an utterly different set of inferential techniques to analyze the large and complex dependence structure among the test statistics, and then derive the normal approximations for the global null distribution and total false discoveries. In summary, our proposal addresses a particularly important class of scientific questions, and makes a useful addition to the toolbox of longitudinal data modeling. 

We adopt the following notations throughout this article. For a vector $a=(a_1,\ldots,a_p)\trans$, let $[a]_j$ denote its $j$th entry and $\|a\|_q=(\sum_{j=1}^p |a_j|^q)^{1/q}$. For a matrix $A$, let $[A]_{j_1, j_2}$ denote its $(j_1,j_2)$th entry. Let $\lambda_{\max}(A)$ and $\lambda_{\min}(A)$ denote the maximum and minimum eigenvalues of $A$, respectively. Let $\|A\|_{\max}=\max_{j,k}|A_{jk}|$, $\|A\|_{1,1}=\sum_{j,k}|A_{jk}|$, and $\|A\|_q=\sup_{\|a\|_q=1}\|A a\|_q$ for $q \geq 1$. Let $\|A\|_0$ denote the number of nonzero entries in $A$, and let $\diag(A)$ denote the diagonal matrix with  its diagonal equal to the diagonal entries of $A$. 
For two positive sequences $a_n$ and $b_n$, $a_n \lesssim b_n$ means that there exists a constant $c > 0$, such that $a_n \leq c b_n$ for all $n$; $a_n \asymp b_n $ if $a_n \lesssim b_n$ and $b_n \lesssim a_n$, and $a_n \ll b_n$ if $\limsup_{n\rightarrow\infty} {a_n}/{b_n}=0$. 

The rest of the article is organized as follows. Section \ref{sec:model} presents the model, and our proposed estimation and inference procedures. Section \ref{sec:theory} establishes the theoretical properties. Section \ref{sec:simulations} reports the simulations, and Section \ref{sec:realdata} presents an application to a longitudinal AD study. Section \ref{sec:discussion} concludes the paper, and the Supplementary Material collects all technical proofs and additional simulation results.

\section{Model Estimation and Inference}
\label{sec:model}

\subsection{High-dimensional multi-response GCM}
\label{subsec:model}

Suppose there are $N$ subjects, $R$ response variables, and for subject $i$, there are longitudinal data collected at $T_i$ time points, $i = 1, \ldots, N$. In this article, we only consider the scenario when $T_1 = \cdots = T_N = T$, and leave the scenario when the subjects have varying numbers of observations as future research. Let $y_{i,r,t} \in \Rbb$ denote the $r$th response variable, $g_{i,t} \in \Rbb$ the time variable, $x_i \in \Rbb^{p}$ the time-invariant predictor vector, and $z_{i,t} \in \Rbb^{q}$ the time-variant predictor vector, for subject $i$ at time $t$, $i = 1, \ldots, N, r = 1, \ldots, R, t = 1, \ldots, T$. For our motivating examples, $y_{i,r,t}$ corresponds to the individual brain region or gene, $g_{i,t}$ is the age variable, $x_i$ collects the binary AD status, and other time-invariant covariates such as sex and education level, and $z_{i,t}$ collects the time-variant covariates such as the cognitive scores. We consider the following classical two-level GCM, at level 1, 
\begin{align} \label{eqn:gcm-level1} 
\textrm{Level 1: } \quad & y_{i,r,t} = \beta_{0,i,r} + \beta_{1,i,r}g_{i,t} + \xi_r\trans z_{i,t} + \epsilon_{i,r,t}, 
\end{align}
where $\beta_{0,i,r}, \beta_{1,i,r} \in \Rbb$ are the individual-level initial state and the growth rate of the mean growth curve for subject $i$ at region $r$, respectively, $\xi_r \in \Rbb^q$ is the time-invariant fixed-effect of $z_{i,t}$, and $\epsilon_{i,r,t} \in \Rbb$ is the random error, and at level 2, 
\begin{align} \label{eqn:gcm-level2} 
\begin{split}
\textrm{Level 2: } \quad & \beta_{0,i,r} = \mu_{0,r} + \gamma_{0,r}\trans x_i + \zeta_{0,i,r}, \\
& \beta_{1,i,r} = \mu_{1,r} + \gamma_{1,r}\trans x_i + \zeta_{1,i,r},
\end{split}
\end{align}
where $\mu_{0,r}, \mu_{1,r}$ are the population-level intercepts, and $\gamma_{0,r}, \gamma_{1,r} \in \Rbb^{p}$ are the population-level slopes for the initial state and the growth rate of the mean growth curve of region $r$, respectively, and $\zeta_{0,i,r}, \zeta_{1,i,r} \in \Rbb$ are the random errors. 

We assume the random errors $\epsilon_{i,r,t}, \zeta_{0,i,r}, \zeta_{1,i,r}$ follow a mean zero normal distribution, 
\vspace{-0.01in} 
\begin{align} \label{eqn:error-structure}
\begin{split}
& (\epsilon_{i,1,1}, \ldots, \epsilon_{i,1,T},\ldots,\epsilon_{i,R,1},\ldots,\epsilon_{i,R,T})\trans \sim \textrm{Normal}(0, \Sigma_R \otimes \Sigma_T), \\
& (\zeta_{0,i,r}, \zeta_{1,i,r})\trans \sim \textrm{Normal}(0, \Sigma_{\zeta}),
\end{split}
\end{align}
where $\otimes$ denotes the Kronecker product, $\Sigma_R \in \Rbb^{R \times R}$ captures the spatial correlation among different regions or genes, $\Sigma_T \in \Rbb^{T \times T}$ captures the temporal correlation across different time points, and $\Sigma_{\zeta} \in \Rbb^{2 \times 2}$ captures the random departure of individual subjects from the population-level intercept and slope. Here we introduce the Kronecker structure to simplify the covariance of the random error $\epsilon_{i,r,t}$. Such a structure has been often adopted in the literature \citep[see, e.g.,][]{YinLi2012, LengTang2012, xia2017hypothesis}. To make $\Sigma_R$ and $\Sigma_T$ identifiable, we assume that $\tr(\Sigma_T) = T$, without loss of generality. 
Additionally, we assume that  $\cov(\beta_{d,i,r_1},\epsilon_{i,r_2,t})=0$ for $d=0,1$ and $r_1,r_2=1,\ldots, R$, and $\cov(\zeta_{d_1,i,r_1}, \zeta_{d_2,i,r_2})=0$ for $d_1,d_2=0,1$ and $r_1\neq r_2$, which are commonly imposed in the GCM literature \citep[see, e.g.,][]{hox2005multilevel,bosker2011multilevel}.

Plugging \eqref{eqn:gcm-level2} into \eqref{eqn:gcm-level1}, we obtain that  
\begin{align} \label{eqn:gcm}
\begin{split}
y_{i,r,t} & = \mu_{0,r} + \mu_{1,r} g_{i,t} + \gamma_{0,r}\trans x_i + \gamma_{1,r}\trans g_{i,t} x_i + \xi_r\trans z_{i,t} + \zeta_{0,i,r} + \zeta_{1,i,r} g_{i,t} + \epsilon_{i,r,t} \\
 & = \big( \mu_{0,r}, \mu_{1,r}, \gamma_{0,r}\trans, \gamma_{1,r}\trans, \xi_r\trans \big) \big( 1, g_{i,t}, x_i\trans, g_{i,t} x_i\trans, z_{i,t}\trans \big)\trans + \left\{ \big( \zeta_{0,i,r}, \zeta_{1,i,r} \big)\trans \big(1, g_{i,t} \big) + \epsilon_{i,r,t} \right\}
\end{split}
\end{align}
Write $y_r= (y_{1,r,1}, \ldots, y_{1,r,T}, \ldots, y_{N,r,1}, \ldots, y_{N,r,T})\trans \in \Rbb^{TN}$, ${\beta^{(r)}} = (\eta_r\trans, \xi_r\trans)\trans \in \Rbb^{2p+q+2}$, $\eta_r = (\mu_{0,r}, \mu_{1,r}, \gamma_{0,r}\trans, \gamma_{1,r}\trans)\trans \in \Rbb^{2p+2}$, $\epsilon_r = (\epsilon_{1,r,1}, \ldots, \epsilon_{1,r,T}, \ldots, \epsilon_{N,r,1}, \ldots, \epsilon_{N,r,T})\trans \in \Rbb^{TN}$, $\zeta_r = (\zeta_{0,1,r}, \zeta_{1,1,r}, \ldots, \zeta_{0,N,r}, \zeta_{1,N,r})\trans \in \Rbb^{2N}$, $r = 1, \ldots, R$, and 
\begin{align*}
\begin{split}
& X = \begin{pmatrix}
X_1 \\
\vdots\\
X_N
\end{pmatrix} \in \Rbb^{TN \times (2p+q+2)}, \quad\quad
G = \diag\left( \{{G}_i\}_{i=1}^N \right) = \begin{pmatrix}
G_1 & \ldots & 0 \\
0 & \ddots & 0 \\
0 & \ldots & G_N 
\end{pmatrix} \in \Rbb^{TN \times 2N}, \\ 
& X_i = \begin{pmatrix}
1 & g_{i,1} & x_i\trans & g_{i,1}x_i\trans & z_{i,1}\trans \\
\vdots & \vdots & \vdots & \vdots & \vdots \\
1 & g_{i,T} & x_i\trans & g_{i,T}x_i\trans & z_{i,T}\trans 
\end{pmatrix} \in \Rbb^{T \times (2p+q+2)}, \;
G_i=\begin{pmatrix}
1 & g_{i,1}   \\
\vdots & \vdots  \\
1 & g_{i,T} 
\end{pmatrix} \in \Rbb^{T \times 2}, \; 
i=1,\ldots,N.
\end{split}
\end{align*}
Then, Model \eqref{eqn:gcm} can be written in the matrix form as,
\begin{align} \label{eqn:model-matrix-form}
(y_1,\ldots,y_R) = X \left( {\beta^{(1)}, \ldots, \beta^{(R)}} \right) + \left( G \zeta_1 + \epsilon_1,\ldots,G \zeta_R+ \epsilon_R \right),
\end{align}
and the covariance matrix of $G \zeta_r+ \epsilon_r$ is of the form, 
\vspace{-0.01in}
\begin{align} \label{eqn:Sigma-rr}
\Sigma^{(r)} = G( I_N \otimes {\Sigma_\zeta} )G\trans + \diag\left( \{[\Sigma_{R}]_{r,r}\Sigma_T\}_{i=1}^N \right) \;\; \in \; \Rbb^{TN \times TN}, \quad r=1, \ldots, R,
\end{align}
where $I_N \in \Rbb^{N \times N}$ is the identity matrix, and $\diag\left( \{[\Sigma_{R}]_{r,r}\Sigma_T\}_{i=1}^N \right)$ is a block-diagonal matrix with all the blocks equal to the same matrix $[\Sigma_{R}]_{r,r}\Sigma_T \in \Rbb^{T \times T}$.

Our goal is to estimate {$\beta^{(r)}$} and $\Sigma^{(r)}$, and then infer the parameters regarding the mean growth curves. Specifically, our inference aims at the population-level intercepts and slopes of the initial state and the growth rate of the mean growth curve of region $r$, i.e., $\eta_r = (\mu_{0,r}, \mu_{1,r}, \gamma_{0,r}\trans, \gamma_{1,r}\trans)\trans$. We study the global test and see if the population-level growth curves are mean zero and unaffected by the predictors $x_i$ for all response variables $y_r$'s, 
\vspace{-0.01in}
\begin{align}\label{eqn:global_test}
H_{0}: (\eta_1, \ldots,\eta_R)=0 \;\; \text{ versus } \;\; H_{1}: (\eta_1, \ldots,\eta_R) \neq 0.
\end{align}
We also study multiple individual tests and aim to identify the nonzero population mean effects as well as the subset of predictors $x_i$ that affect some response variable $y_r$,  \begin{align} \label{eqn:multiple_test}
H_{0,r,j}: [\eta_{r}]_{j} = 0 \;\; \text{ versus } \;\; H_{1,r,j}: [\eta_{r}]_{j} \neq 0, \quad r=1,\ldots,R, \; j=1,\ldots,2p+2.
\end{align}
Meanwhile, we control the false discovery rate (FDR) and the false discovery proportion (FDP) at a pre-specified level. We remark that no existing literature simultaneously tackles the inference problems \eqref{eqn:global_test} and \eqref{eqn:multiple_test} in a high-dimensional multi-response GCM setting.

\subsection{Parameter estimation}
\label{subsec:estimation}

We recognize the key challenge of parameter estimation for our model is the covariance matrix $\Sigma^{(r)}$ in \eqref{eqn:Sigma-rr}, for $r = 1, \ldots, R$, as it involves a large number of unknown parameters when the dimension of the response $R$ is large, while the number of subjects $N$ and the number of time points $T$ are often limited. In addition, it involves three covariance matrices, $\Sigma_R, \Sigma_T, \Sigma_\zeta$, which need to be decoupled from each other and be estimated separately. We next develop a novel five-step procedure to estimate $\Sigma^{(r)}$. 

In Step 1, we first estimate the off-diagonal elements of the spatial covariance matrix $\Sigma_R$. We comment that only the diagonal elements of $\Sigma_R$ are required in $\Sigma^{(r)}$ in \eqref{eqn:Sigma-rr}, but the off-diagonal elements of $\Sigma_R$ turn out to be useful for the estimation of $\Sigma_T, \Sigma_\zeta$, and are also easier to estimate than the diagonal elements. Note that, for each individual $i$, the spatial variance averaging over the time points can be written as, 
\begin{align*}
\Sigma_{1,i} = \frac{1}{T} \sum_{t=1}^{T}\var\left\{ (y_{i,1,t}, \ldots, y_{i, R, t})\trans \right\} = \frac{\tr(G_i \Sigma_\zeta G_i\trans)}{T}I_R + \Sigma_R, 
\end{align*}
for $i = 1, \ldots, N$. A sample variance estimator of $\Sigma_{1,i}$ is, 
\begin{align} \label{eqn:SigmaR-offdiag1}
\hat{\Sigma}_{1,i} = \frac{1}{T} \sum_{t=1}^{T} (\check{y}_{i,1,t}, \ldots, \check{y}_{i, R, t})\trans (\check{y}_{i,1,t}, \ldots, \check{y}_{i, R, t}) \in \Rbb^{R \times R}. 
\end{align}
where $\check{y}_{i,r,t}=y_{i,r,t} - \bar{y}_{r,t}$ is the centered response across subjects with $\bar{y}_{r,t} = N^{-1} \sum_{i=1}^N y_{i,r,t}$. Here the centering is with respect to subjects, as the subjects are independent, but different regions and time points are not. We use the same centering for other sample variance and covariance estimators later, and we show they achieve the desirable convergence rates. We also note that the first term in ${\Sigma}_{1,i}$ is a diagonal matrix, and thus we can pool both individual and time information to estimate the off-diagonal elements of $\Sigma_R$ as,
\begin{align} \label{eqn:sigmaR-offdiag2}
[\hat{\Sigma}_R]_{r_1,r_2} = \left[ \hat{\Sigma}_{1} - \text{diag}(\hat{\Sigma}_{1}) \right]_{r_1,r_2}, \text{ with }\hat{\Sigma}_{1} = \frac{1}{N} \sum_{i=1}^N \hat{\Sigma}_{1,i}, 
\;\; r_1, r_2 = 1, \ldots, R, r_1 \neq r_2. 
\end{align}

In Step 2, we estimate the temporal covariance matrix $\Sigma_T$. Note that, for each individual $i$, the temporal covariance between two regions can be written as, 
\begin{align*}
\Sigma_{2, r_1, r_2} = \cov\left\{ (y_{i,r_1,1}, \ldots, y_{i,r_1,T})\trans, (y_{i,r_2,1}, \ldots, y_{i,r_2,T})\trans \right\} = [\Sigma_{R}]_{r_1, r_2} \Sigma_T, 
\end{align*}
for $i = 1, \ldots, N, r_1, r_2 = 1, \ldots, R, r_1 \neq r_2$. A sample covariance estimator of $\Sigma_{2, r_1, r_2}$ is, 
\begin{align} \label{eqn:SigmaT1}
{\hat{\Sigma}_{2, r_1, r_2} = \frac{1}{N} \sum_{i=1}^N 
(\check y_{i,r_1,1} , \ldots, \check y_{i,r_1,T})\trans (\check y_{i,r_2,1} , \ldots, \check y_{i,r_2,T})} \in \Rbb^{T \times T},
\end{align}
where $\check{y}_{i,r,t}$ is as defined before. Therefore, we can pool both individual and region information to estimate $\Sigma_T$, based on the average of $\hat{\Sigma}_{2, r_1, r_2}$ across the pairs $(r_1, r_2)$, along with the estimator $[\hat{\Sigma}_R]_{r_1,r_2}$ from Step 1. In addition, to help achieve a desired convergence rate for our estimator of $\Sigma_T$, we propose to choose the set of pairs $(r_1, r_2)$ with the largest $K$ off-diagonal entries $[\hat{\Sigma}_R]_{r_1,r_2}$ among all $r_1, r_2 = 1, \ldots, R, r_1 < r_2$, and we denote this set as $\mathcal S$. Our study suggests that, as long as $K$ is of the same order of $R$, the corresponding estimator has a desired convergence rate, and thus we set $K = R$ in our implementation. That is, we estimate the temporal covariance matrix $\Sigma_T$ as,
\begin{align} \label{eqn:SigmaT2}
\hat{\Sigma}_T= \frac{1}{R [\hat{\Sigma}_R]_{r_1,r_2}} \sum_{(r_1,r_2) \in \mathcal S} \hat{\Sigma}_{2,r_1,r_2}. 
\end{align}

In Step 3, we estimate the covariance matrix $\Sigma_\zeta$. Note that, for each individual $i$, the temporal variance averaging over regions can be written as, 
\begin{align*}
\Sigma_{3,i} = \frac{1}{R} \sum_{r=1}^{R}\var\left\{ (y_{i,r,1}, \ldots, y_{i, r, T})\trans \right\} = G_i \Sigma_\zeta G_i\trans+ \frac{\tr(\Sigma_R)}{R}\Sigma_T,
\end{align*}
for $i=1,\ldots,N$. A sample variance estimator of $\Sigma_{3,i}$ is, 
\begin{align} \label{eqn:SigmaPsi1}
{\hat{\Sigma}_{3,i} = \frac{1}{R} \sum_{r=1}^{R} (\check{y}_{i,r,1}, \ldots, \check{y}_{i, r, T})\trans (\check{y}_{i,r,1}, \ldots, \check{y}_{i, r, T})} \in \Rbb^{T \times T}. 
\end{align}
Thus, to estimate $\Sigma_\zeta$, we need to estimate $\kappa = \tr(\Sigma_R) / R$, and plug in the estimate of $\Sigma_T$ from Step 2. Note that, as long as the number of time points $T \ge 3$, there exists a vector $u_i \in \Rbb^T$, such that $\|u_i\|_2=1$, and $u_i\trans G_i=(0,0)$, for $i=1,\ldots,N$. Correspondingly, $u_i\trans {\Sigma}_{3,i} u_i = \kappa (u_i\trans {\Sigma}_{T} u_i)$. Therefore, we estimate $\kappa$ as, 
\begin{equation} \label{eqn:kappa}
\hat \kappa=\frac{\sum_{i=1}^N u_i\trans {\hat\Sigma}_{3,i} u_i}{\sum_{i=1}^N u_i\trans {\hat\Sigma}_{T} u_i}.
\end{equation}
Similarly, as long as the two columns of $G_i$ are linearly independent for $i=1,\ldots,N$, there exist vectors $v_{i,1}, v_{i,2} \in \Rbb^T$, such that $v_{i,1}\trans G_i=(1,0)$, and $v_{i,2}\trans G_i=(0,1)$. Correspondingly, $v_{i,j_1}\trans \Sigma_{3,i} v_{i,j_2} = [\Sigma_\zeta]_{j_1,j_2} + \kappa (v_{i,j_1}\trans \Sigma_{T} v_{i,j_2})$, $j_1, j_2 = 1, 2$. Therefore, we estimate $\Sigma_\zeta$ as, 
\begin{equation} \label{eq:SigmaPsi2}
[\hat{\Sigma}_\zeta]_{j_1,j_2} = \frac{1}{N} \sum_{i=1}^N \left\{ \left( v_{i,j_1}\trans {\hat\Sigma}_{3,i} v_{i,j_2} \right) - \hat\kappa \left( v_{i,j_1}\trans {\hat\Sigma}_{T} v_{i,j_2} \right) \right\}, \quad j_1, j_2 = 1, 2.
\end{equation}

In Step 4, we estimate the diagonal elements of the spatial covariance matrix $\Sigma_R$. Recall $\tr(\Sigma_{1,i}) = \tr(G_i \Sigma_\zeta G_i\trans) R / T + \tr(\Sigma_R)$. 
Therefore, we can estimate $(NT)^{-1} \sum_{i=1}^N \tr(G_i\Psi G_i\trans)$ by $\tr(\hat\Sigma_{1})/R - \hat \kappa$. Correspondingly, we estimate the diagonal elements of $\Sigma_R$ as 
\begin{equation} \label{eqn:SigmaR-diag}
[\hat\Sigma_{R}]_{r,r} = \left[\text{diag}(\hat\Sigma_{1}) - \text{diag}\left( \tr(\hat\Sigma_{1})/R - \hat \kappa \right) \right]_{r,r}, \quad r = 1, \ldots, R.
\end{equation}

Finally, in Step 5, we plug the estimates $\hat\Sigma_T$, $\hat\Sigma_\zeta$, and $[\hat\Sigma_{R}]_{r,r}$ into \eqref{eqn:Sigma-rr} to obtain an estimate $\hat\Sigma^{(r)}$ of $\Sigma^{(r)}$. We summarize our estimation procedure in Algorithm \ref{alg1}.

\begin{algorithm}[t!]
\caption{Covariance estimation procedure}
\label{alg1}
\begin{enumerate}[label=Step \arabic*.,leftmargin=*]
\item Estimate the off-diagonal elements of the spatial covariance matrix $\Sigma_R$ via \eqref{eqn:sigmaR-offdiag2}. 

\item Estimate the temporal covariance matrix $\Sigma_T$ via \eqref{eqn:SigmaT2}.

\item Estimate the covariance matrix $\Sigma_\zeta$ via \eqref{eq:SigmaPsi2}. 

\item Estimate the diagonal elements of the spatial covariance matrix $\Sigma_R$ via \eqref{eqn:SigmaR-diag}. 

\item Estimate $\Sigma^{(r)}$ by plugging the estimates $\hat\Sigma_T$, $\hat\Sigma_\zeta$, and $[\hat\Sigma_{R}]_{r,r}$ into \eqref{eqn:Sigma-rr}. 
\end{enumerate} 
\end{algorithm}

Once obtaining $\hat\Sigma^{(r)}$, we estimate $\beta^{(r)}$ via least-squares straightforwardly as,
\begin{align} \label{eqn:beta}
\hat{\beta}^{(r)} = \left\{ X\trans \left( \hat\Sigma^{(r)} \right)^{-1} X \right\}^{-1} X\trans \left( \hat\Sigma^{(r)} \right)^{-1} y_r, \quad r=1, \ldots, R.
\end{align}

We make a few remarks regarding our estimation procedure. 

First, we note that, to obtain a good estimator for $\Sigma^{(r)}$, we need to respectively estimate $\Sigma_R, \Sigma_T$, and $\Sigma_\zeta$. There are different options to estimate $\Sigma_T$. One is to pool $NR$ correlated samples together, e.g., through averaging over individual $\hat{\Sigma}_{3,i}$ in \eqref{eqn:SigmaPsi1}. However, $\Sigma_T$ and $\Sigma_\zeta$ are not easily separable if we go this way. Recognizing that the random departures are uncorrelated with each other across regions, we take another option, by first obtaining the time covariance across different regions, i.e., $[\Sigma_{R}]_{r_1, r_2} \Sigma_T$, then pooling $TN$ correlated samples to estimate the off-diagonal elements of $\Sigma_R$, then averaging over individuals and different region pairs to estimate $\Sigma_T$. Subsequently, we pool $NR$ samples to estimate $\Sigma_\zeta$, and pool $NT$ samples to estimate $[\Sigma_{R}]_{r,r}$. Such sample pooling steps are crucial to ensure an accurate estimation of the covariance matrices $\Sigma_R, \Sigma_T$, and $\Sigma_\zeta$. 

Second, we estimate $\Sigma_{1, i}$ and $\Sigma_{3, i}$ via \eqref{eqn:SigmaR-offdiag1} and \eqref{eqn:SigmaPsi1}, respectively. However, we note that, $\Ebb(y_{i,r,t}) = [X]_{(i-1)T+t,\cdot} {\beta^{(r)}}$, where $[X]_{(i-1)T+t, \cdot}$ is the $\{(i-1)T+t\}$th row of $X$, and thus the means of $y_{i,r,t}$ are different for all $(i,r,t), i=1,\ldots, N, r=1, \ldots, R, t=1, \ldots, T$. Consequently, both $\hat\Sigma_{1, i}$ and $\hat\Sigma_{3, i}$ are not the usual sample covariance matrices. We later develop a set of new tools to establish their convergence rates in Theorem \ref{lm:rate}.

Last but not least, in our motivating applications, the number of time-invariant and time-variant predictors $(p,q)$ are both relatively small compared to the product of the sample size and the number of time points $NT$. As such, we simply estimate {$\beta^{(r)}$} using the least squares \eqref{eqn:beta}. Meanwhile, our method can be extended to more general scenarios in a straightforward fashion. For instance, if $\max(p,q)$ is large, we can apply the debiased Lasso estimator \citep{zhang2014confidence} by imposing certain sparsity structures on {$\beta^{(r)}$'s}.

\subsection{Hypothesis testing}
\label{subsec:inference}

We next develop a global testing procedure for the hypothesis in \eqref{eqn:global_test}, then a multiple testing procedure for the hypotheses in \eqref{eqn:multiple_test}.

For global testing, we consider the following test statistic,
\begin{align} \label{eqn:global_test_stat}
J = \max_{r=1,\ldots,R, j=1, \ldots, {2p+2}} J_{r,j}^2, \quad \textrm{ where } \;  J_{r,j}^2=\frac{\left( \left[ \hat{\beta}^{(r)} \right]_j \right)^2}{\left[ \left\{ X\trans \left( \hat\Sigma^{(r)} \right)^{-1} X \right\}^{-1}\right]_{j,j}},
\end{align}
We then propose a global testing procedure as summarized in Algorithm \ref{alg_global}. The test is built on the asymptotic property of the test statistic $J$, as we establish in Section \ref{sec:theory}. Intuitively, with some mild dependence conditions, $\{J_{r,j}: r=1,\ldots,R, j =1, \ldots, 2p+2\}$ are close to weakly dependent standard normal random variables under $H_{0}$. Therefore, the proposed test statistic $J$, which takes the form of the maximum of the square of $J_{r,j}$'s, should be close to $2 \log \tilde{p}$ under the null hypothesis, where $\tilde{p} = (2p+2)R$.
 
\begin{algorithm}[t!]
\caption{Global testing procedure}
\label{alg_global}
\begin{enumerate}[label=Step \arabic*.,leftmargin=*]
\item Compute the test statistics via \eqref{eqn:global_test_stat}. 

\item Define the global test, 
\begin{align*}
\Psi_\alpha = I\left( J \ge 2 \log \tilde{p} -  \log \log \tilde{p} + q_{\alpha} \right), 
\end{align*}
where $q_\alpha = - \log \pi  - 2 \log\{\log(1-\alpha)^{-1}\}$, $I(\cdot)$ is the indicator function, and $\alpha$ is the pre-specified significance level. 

\item Reject the null hypothesis $H_0$ in \eqref{eqn:global_test} if $\Psi_\alpha=1$.
\end{enumerate} 
\end{algorithm}

\begin{algorithm}[b!]
\caption{Multiple testing procedure}
\label{alg_multiple}
\begin{enumerate}[label=Step \arabic*.,leftmargin=*]
\item Compute the individual test statistics $J_{r,j}$ via \eqref{eqn:global_test_stat}, for $r=1,\ldots,R, j=1,\ldots,2p+2$. 

\item Estimate the FDP by 
\begin{align*}
\widehat{\text{FDP} }(\tau)= \frac{2\{1-\Phi(\tau)\}\tilde{p}}{\sum_{(r,j) \in \Hcal} I(|J_{r,j}|>\tau) \vee 1}.
\end{align*}

\item Compute the threshold value 
\begin{align*}
\hat\tau = \inf \left\{ 0 \le \tau \le t_{\tilde{p}}: \widehat{\text{FDP} }(\tau) \le \alpha \right\}, \;\; \text{ where } \;\; t_{\tilde{p}} = \left( 2 \log \tilde{p} - 2 \log \log \tilde{p} \right)^{1/2}.
\end{align*}
If $\hat\tau$ dos not exist, set $\hat\tau = (2 \log \tilde{p})^{1/2}$.

\item Reject $H_{0,r,j}$ if $|J_{r,j}| \ge \hat\tau$, for $r=1,\ldots,R, j=1,\ldots,2p+2$.
\end{enumerate}
\end{algorithm}

For multiple testing, the key is to control the false discovery, and we propose a multiple testing procedure as summarized in Algorithm \ref{alg_multiple}. Let $\tau$ denote the threshold value such that $H_{0,r,j}$ is rejected if $|J_{r,j}| \geq \tau$, $r=1, \ldots, R, j=1,\ldots, 2p+2$. Let $\Hcal_0 = \{(r,j): \beta^{(r)}_{j}=0, r=1, \ldots, R, j=1,\ldots, 2p+2\}$ denote the set of true null hypotheses and let $\Hcal = \{(r,j): r=1, \ldots, R, j=1,\ldots, 2p+2\}$. Then the FDP and the FDR are, respectively, 
\begin{align*}
\text{FDP}(\tau)=\frac{\sum_{(r,j) \in \Hcal_0} I(|J_{r,j}|>\tau)}{\sum_{(r,j) \in \Hcal} I(|J_{r,j}|>\tau) \vee 1}, \quad\quad
\text{FDR}=\Ebb \{\text{FDP}(\tau)\}.
\end{align*}
We aim to find a threshold $\tau$ so that we can reject as many true positives as possible while controlling the estimated FDP at the pre-specified level $\alpha$. We note that, since the set of true nulls $\Hcal_0$ is unknown, we estimate $|\Hcal_0|$ by $\tilde p$ under the belief that the nulls are dominant among the tests. Subsequently, we estimate the numerator in FDP by $2\{1-\Phi(\tau)\}\tilde{p}$ based on normal approximation. We also note that, the number of false discoveries may not be appropriately estimated if $\hat\tau$ exceeds a certain threshold \citep{xia2018multiple}. We thus set a range $[0,t_{\tilde p}]$ for selecting $\hat\tau$, and threshold it at $(2 \log \tilde{p})^{1/2}$ if it is not attained in the range.

\section{Theoretical Properties}
\label{sec:theory}

\subsection{Regularity conditions}

We now study the theoretical properties of the proposed estimation and inference methods. We begin with a set of regularity conditions. 

\begin{enumerate}[label=(C\arabic*), series=C]
\item \label{C1} 
Suppose $c_1^{-1} \le (NT)^{-1} \lambda_{\min}(X\trans X) \le (NT)^{-1} \lambda_{\max}(X\trans X) \le c_1$, $\|G\|_{\max} \le c_1$, and {$\min_{1\leq i \leq N}\lambda_{\min}(G_i\trans G_i) \ge c_1 ^{-1}$}, for some constant $c_1 > 0$. 

\item \label{C2} 
Suppose $c_2^{-1} \le \lambda_{\min}(\Sigma_T) \le \lambda_{\max}(\Sigma_T) \le c_2$, $c_2^{-1} \le \lambda_{\min}(\Sigma_R) \le \lambda_{\max}(\Sigma_R) \le c_2$, and $T \lambda_{\max}(\Sigma_\zeta) \leq c_2$,  for some constant $c_2>0$.
	
\item \label{C3} 
Suppose there exist at least $R$ entries of $[\Sigma_{R}]_{r_1, r_2}$, $r_1<r_2$, such that $\left| [\Sigma_{R}]_{r_1, r_2} \right| \ge c_3$ for some constant $c_3>0$.
	
\item \label{C5} 
Denote $\Sigma^{(r_1,r_2)}=\diag\left( \{[\Sigma_{R}]_{r_1,r_2} \Sigma_T\}_{i=1}^N \right)$, for $r_1, r_2 = 1, \ldots, R, r_1\neq r_2$, {$\tilde\Sigma^{(r_1,r_2)} = \{X\trans (\Sigma^{(r_1,r_1)})^{-1}X\}^{-1} X\trans (\Sigma^{(r_1,r_1)})^{-1} \Sigma^{(r_1,r_2)} (\{\Sigma^{(r_2,r_2)}\}^{-1}X)^{-1} \{ X\trans (\Sigma^{(r_2,r_2)})^{-1} X \}^{-1}$}, and $D_r = \diag\{(X\trans (\Sigma^{(r)})^{-1} X)^{-1}\}$, for $r=1,\ldots,R$. Denote $\check{\Sigma}^{(r_1,r_2)} = D_{r_1}^{-1/2} \tilde\Sigma^{(r_1,r_2)}  D_{r_2}^{-1/2}$. Suppose $\max_{r_1,r_2} \max_{j_1 < j_2} \left| [\check{\Sigma}^{(r_1,r_2)}]_{j_1,j_2} \right| \le c_4$, and $\max_{r_1\neq r_2}\max_{j_1=j_2} \left| [\check{\Sigma}^{(r_1,r_2)}]_{j_1,j_2} \right| \le c_4$, for some constant $0<c_4<1$.   

\item \label{C6} 
Denote $\beta^{(r)}_{-1} \in \Rbb^{2p+q+1}$ as the sub-vector of $\beta^{(r)}$ with the first entry $\mu_{0,r}$ removed, and $B = \left( \beta^{(1)}_{-1}, \ldots, \beta^{(R)}_{-1} \right) \in \Rbb^{(2p+q+1) \times R}$. Let $s_{B}=\|B\|_0$, $c_{B} = \|B\|_{\max}$, and $c_R = \max_{1 \le r \le R} \| \beta^{(r)}_{-1} \|_2^2$. Let $\theta_{N,T,R,B} = T[\{\log R/(NT)\}^{1/2} + \{\log T/(NR)\}^{1/2}+c_R+{ s_B c_B^2} TR^{-1}]$. Suppose $T (\log \tilde p) \ \theta_{N,T,R,B}=o(1)$.
\end{enumerate}

\noindent
Condition \ref{C1} imposes some mild bounded eigenvalue requirement on the design matrix $X$. In this article, we treat $X$ as deterministic, and similar conditions have been commonly imposed; see, e.g., \cite{zhang2014confidence}. When the design matrix $X$ is random, we can simply replace this condition with the bounded eigenvalue requirement on the covariance of $X$. Moreover, the condition on $\{G_i\trans G_i, i=1,\ldots,N\}$ is mild, as each of them is a $2 \times 2$ matrix. Condition \ref{C2}  is placed on the eigenvalues of $\Sigma_R$ and $\Sigma_T$, which is relatively mild too, as it requires that most of the variables are not highly correlated with each other across regions or over the time points. In addition, the condition on $T \lambda_{\max}(\Sigma_\zeta)$ ensures the bounded eigenvalue property of the covariance matrix $\Sigma^{(r)}$. Condition \ref{C3} naturally holds for a general region-wise dependence structure, and it is only a sufficient condition for Theorem \ref{lm:rate}. Condition \ref{C5} requires the correlations are bounded away from $-1$ and $1$, and basically  excludes the singular cases. Moreover, Conditions \ref{C2} and \ref{C5}  are commonly assumed in the high-dimensional literature \citep[e.g.,][]{bickel2008regularized, yuan2010high, cai2014two,liu2013ggm,xia2015testing}. Condition \ref{C6} regulates the relation of $T$, $R$, $N$, $p$, which is satisfied in our motivating example. It also imposes some level of sparsity on $B$, in that it requires the number of nonzero entries in $B$ and their maximum magnitude are not too large. This condition is reasonable in the motivating examples, and it characterizes the mean variations across subjects, regions and time points. Recall that our covariance estimators in \eqref{eqn:SigmaR-offdiag1}, \eqref{eqn:SigmaT1} and \eqref{eqn:SigmaPsi1} are not the usual sample covariances, because the means of the response variables are different for different subjects, regions and time points. As such, Condition \ref{C6} ensures that our covariance estimators perform asymptotically the same as the sample covariance estimators obtained by the i.i.d. centralized error terms. On the other hand, we do not impose any particular sparsity structures on the coefficient matrix $B$ nor the covariances $\Sigma_R$, $\Sigma_T$ and $\Sigma_\zeta$. As such, we do not employ any penalized procedure such as Lasso in our parameter estimation.

\subsection{False discovery control and power}

First, we establish the convergence rate of the covariance matrix estimator $\hat\Sigma^{(r)}$ from Algorithm \ref{alg1}, which is key for the subsequent size, power and error rate control analyses. 

\begin{theorem} \label{lm:rate}
Suppose Conditions \ref{C1} to \ref{C3} hold, and $\theta_{N,T,R,B}=o(1)$. Then,
\begin{align*}
\| \hat\Sigma^{(r)} - \Sigma^{(r)} \|_{\max} = O_P(\theta_{N,T,R,B}).
\end{align*}
\end{theorem}

\noindent
We remark that, because the means of $y_{i,r,t}$ are different for all $\{(i,r,t), i=1,\ldots N, r=1, \ldots, R, t=1, \ldots, T\}$, the asymptotics of the classical sample covariance estimator based on i.i.d.\ observations are not directly applicable to the sample covariance estimators such as $\hat\Sigma_{1, i}$, $\hat\Sigma_{2,r_1,r_2}$ and $\hat\Sigma_{3, i}$ in \eqref{eqn:SigmaR-offdiag1}, \eqref{eqn:SigmaT1} and \eqref{eqn:SigmaPsi1}. To obtain the desired convergence rate, we first show in Step 1 of the proof of Theorem \ref{lm:rate} the mean variations of the response variables are negligible across different $\{(i,r,t)\}$. We then repeatedly apply an advanced version of Hanson-Wright inequality \citep{chen2021testing} to derive the convergence rates of $[\hat\Sigma_{R}]_{r_1,r_2}$, $\hat\Sigma_T$, $\hat\Sigma_\zeta$, and $[\hat\Sigma_{R}]_{r,r}$ in turn, which eventually leads to the convergence rate of $\hat\Sigma^{(r)}$. 

Next, we derive the limiting distribution of the test statistic $J$ in \eqref{eqn:global_test_stat} under the null hypothesis of the global test \eqref{eqn:global_test}. We show that, $(J- 2 \log \tilde{p}+  \log \log \tilde{p})$ weakly converges to a Gumbel random variable with the cumulative distribution function, $\exp\{-\pi^{-1/2} \exp(-\phi/2)\}$, under the null. We establish the asymptotics when both the product of the number of subjects and the number of time points $NT$, and the product of the number of responses and the covariates $\tilde{p} = (2p+2)R$ diverge to infinity.
\begin{theorem} \label{thm:data_global_testing}
Suppose Conditions \ref{C1} to \ref{C6} hold. Then  for any $\phi \in \Rbb$,
\begin{align*}
\Pbb_{H_0}\left( J - 2 \log \tilde{p}+  \log \log \tilde{p} \le \phi \right) \to \exp\left\{ -\pi^{-1/2} \exp(-{\phi}/{2}) \right\}, \text{ as } NT \textrm{ and } \tilde{p} \to \infty.
\end{align*}
\end{theorem}

Next, we study the asymptotic power of the proposed global test in Algorithm \ref{alg_global}. We consider the following class of regression coefficients $\{\beta^{(r)}, r=1,\ldots, R\}$,
\vspace{-0.01in}
\begin{align*}
\Ucal(c) = \left\{\{\beta^{(r)}_{j}\}_{(r,j) \in \Hcal} : \max_{r,j}\frac{|\beta^{(r)}_{j}|}{\left[\{ X\trans (\Sigma^{(r)})^{-1} X \}^{-1}\right]_{j,j}^{1/2}} \ge c (\log \tilde{p})^{1/2}\right\}.
\end{align*}
Since $\left[\{ X\trans (\Sigma^{(r)})^{-1} X \}^{-1}\right]_{j,j}^{1/2}$ is of order $(NT)^{-1/2}$, the above class includes all coefficient matrices that have one of the entries with a magnitude of the order $\{\log \tilde{p}/(NT)\}^{1/2}$.

\begin{theorem} \label{thm:data_multiple_power}
Suppose Conditions \ref{C1} to \ref{C3}, and {\ref{C6}} hold. Then, 
\begin{align*}
\inf_{\{\beta^{(r)}_{j}\} \in \Ucal(2\sqrt{2})}\Pbb(\Psi_{\alpha}=1) \to 1, \text{ as } NT \textrm{ and } \tilde{p} \to \infty.
\end{align*}
\end{theorem}

\noindent
We immediately see that, based on this theorem, if we set the constant $c$ as $2\sqrt{2}$, then our proposed global test enjoys a full power asymptotically. 

Finally, we establish the error rate control of the proposed multiple testing procedure in Algorithm \ref{alg_multiple}. We introduce one more regularity condition \ref{C7}, which ensures that $\hat\tau$ is attained in the range $[0,(2 \log \tilde{p} - 2 \log \log \tilde{p})^{1/2}]$. We note that Condition \ref{C7} only requires a few entries of $\{\beta^{(r)}_{j}\}$ to have a  magnitude of the order $\{\log \tilde{p}\}^{(1+\rho)/2}/(NT)^{1/2}$, and is thus mild. Moreover, when this condition is not satisfied, the asymptotic FDR control can still be obtained but more conservatively.

\begin{enumerate}[resume*=C]
\item \label{C7}
Denote $S_{\rho} = \left\{(r,j) \in \Hcal: {\left\{\beta^{(r)}_{j}\right\}^2} / \left[\{ X\trans (\Sigma^{(r)})^{-1} X \}^{-1}\right]_{j,j} \ge (\log \tilde{p})^{1+\rho} \right\}$. Suppose $|S_{\rho}| \ge \{1/(\pi^{1/2}\alpha)+\delta\}(\log \tilde{p})^{1/2}$ for some $\rho>0$ and $\delta > 0$. 
\end{enumerate}	

\begin{theorem} \label{thm:data_FDR_FDP}
Suppose Conditions \ref{C1} to \ref{C7} hold, and $\tilde{p}_0=|\Hcal_0| \asymp \tilde{p}$. Then, 
\begin{align*}
\lim_{(NT, \tilde{p}) \to \infty} \frac{FDR}{\alpha \tilde{p}_0/\tilde{p}} = 1, \lim_{(NT, \tilde{p}) \to \infty} \frac{FDP(\hat\tau)}{\alpha \tilde{p}_0/\tilde{p}} = 1 \text{ in probability.}
\end{align*}
\end{theorem}

\subsection{Extension to sub-Gaussian distribution}
\label{subsec:subGaussian}

In our GCM model, we have assumed that the errors follow a Gaussian distribution as in \eqref{eqn:error-structure}. We now extend it to the sub-Gaussian distribution. More specifically, instead of assuming that $\Sigma_T^{-1/2} \epsilon_i \Sigma_R^{-1/2}$ and $\Sigma_\zeta^{-1/2} \zeta_{i,r}$ have i.i.d.\ Gaussian entries, we assume they have i.i.d.\ sub-Gaussian entries, and study the corresponding theoretical properties. We begin with a regularity condition. 

\begin{enumerate}[resume*=C]
\item \label{C8}
Suppose $\log\tilde{p} = o(\{N/ \max(p,q)\}^{1/4})$, {$\| X \|_{\max} = O(1)$}, and 
\begin{align*}
\Ebb\{\exp(\nu\epsilon_{i,r,t}^2)\} \leq c_5, \quad \text{ and } \quad \Ebb\{\exp(\nu\zeta_{d,i,r}^2) \} \leq c_5, 
\end{align*}
for some constants $\nu, c_5 > 0$, $i=1,\ldots,N$, $r=1, \ldots, R$, $t=1,\ldots,T$, and $d=0,1$.
\end{enumerate}

\noindent
Condition \ref{C8} is mild, as both $p$ and $q$ are small in our targeted settings, and the bounded design matrix is also reasonable. 

Next, we establish the theoretical properties under the sub-Gaussian scenario. 

\begin{theorem} \label{thm:subgaussian_data}
Suppose Condition \ref{C8} holds.
\begin{enumerate}[label=(\roman*),leftmargin=*]
\item Suppose the same conditions in Theorem \ref{thm:data_global_testing} hold. Then, for any $\phi \in \Rbb$
\[
\Pbb_{H_0}\left( J - 2 \log \tilde{p}+  \log \log \tilde{p} \le \phi \right) \to \exp\left\{ -\pi^{-1/2} \exp(-{\phi}/{2}) \right\}, 	\text{ as } \tilde{p} \to \infty.
\]

\item Suppose the same conditions in Theorem \ref{thm:data_multiple_power} hold. Then, 
\[
\inf_{\{\beta^{(r)}_{j}\} \in \Ucal(2\sqrt{2})}\Pbb(\Psi_{\alpha}=1) \to 1, 	\text{ as } \tilde{p} \to \infty.
\]

\item Suppose the same conditions in Theorem \ref{thm:data_FDR_FDP} hold, and $(\log \tilde{p})^{7+\varepsilon} = O\{N/\max(p,q)\}$ for some small constant $\varepsilon>0$. Then,  
\[
\lim_{\tilde{p} \to \infty} \frac{FDR}{\alpha \tilde{p}_0/\tilde{p}} = 1, \lim_{\tilde{p} \to \infty} \frac{FDP(\hat\tau)}{\alpha \tilde{p}_0/\tilde{p}} = 1 \text{ in probability.} 
\]
\end{enumerate}
\end{theorem}

\noindent
We remark that, under the sub-Gaussian condition, our test statistic can be written as a sum of $2N+NT$ random variables with unequal variances. In comparison to the well established normal approximation where the test statistic can be expressed as the sum of i.i.d.\ sub-Gaussian random variables, our case is more challenging, as we need to perform truncations on those $2N+NT$ non-identically distributed random variables, which further leads to a more complicated normal approximation on the sum of those variables. See the proof of Theorem \ref{thm:subgaussian_data} for more details.

\section{Simulation Studies}
\label{sec:simulations}

\subsection{Simulation setup}
\label{subsec:sim-setup}

We study the finite-sample performance of the proposed method. We also compare with a simple alternative solution that fits one response variable at a time using the restricted maximum likelihood (REML) approach. That is, we obtain the coefficient estimates using REML, compute the test statistic $J_{r,j}$ following \eqref{eqn:global_test_stat}, then carry out the the global and multiple testing procedures accordingly. The implementation of REML is based on the R packages \texttt{lme4} and \texttt{lmerTest} \citep{Douglas2015Fitting,lmertest}. 

We simulate the model following the setup in \eqref{eqn:model-matrix-form} and \eqref{eqn:Sigma-rr}. We set $N=\{100,200\}$, $T=\{4,8\}$, $R=\{50,100\}$, $p=10$, and $q=2$. We draw $g_{i,t}$ randomly from Uniform$[0,1]$, and generate each entry of $x_i$ and $z_{i,t}$ independently from Normal$(0,1)$, for $i =1, \ldots, N$, $t = 1,\ldots, T$. We randomly set 5\% of the coefficients $\{\xi_r\}$ as non-zero, which equal 0.2 in the global testing case and 0.5 in the estimation evaluations and the multiple testing case. 

We consider two structures for the temporal covariance matrix $\Sigma_T$, an autoregressive structure and a moving average structure. Specifically, we begin with ${[\Sigma^{'}_{T}]_{t_1,t_2}} = 0.4^{|t_1-t_2|}$ for $1 \le t_1,t_2 \le T$, and $[\Sigma^{'}_{T}]_{t_1,t_2}= 1/(|t_1-t_2|+1)$ for $|t_1 - t_2| \le 3$ and $0$ otherwise. We next set $\Sigma^{''}_{T} = \Sigma^{'}_{T} \odot V_T$, to adopt different variances among the time points, where $\odot$ is the Hadamard product, and $V_{T} = u_T u_T\trans$, with $u_T=(1,2,3,4)\trans$ for $T=4$, and $u_T=(1,2,3,4,1,2,3,4)\trans$ for $T=8$. We then set $\Sigma_{T} = \{ T/\tr(\Sigma^{''}_{T}) \} \Sigma^{''}_{T}$, so that $\tr(\Sigma_{T})=T$. We also consider two structures for the spatial covariance matrix $\Sigma_R$. Specifically, we begin with the precision matrix $\Omega^{'}_R$, and consider a hub graph where the nodes are evenly partitioned into disjoint groups with 5 nodes each while there exists one node connecting all the other nodes inside each group, and a small-world graph, with one starting neighbor and 5\% probability of rewiring. {We set the diagonal values to one, and draw the non-zero entries of the off-diagonal values randomly from Uniform$[-0.6,-0.2] \cup [0.2,0.6]$}. We next set $\Omega^{''}_{R} = (\Omega^{'}_{R} + \delta_R I_R) / (1+\delta_R)$, where $\delta_R = \max\{ 0.05, -\lambda_{\min}(\Omega^{'}_{R}) \}$. We then set $\Sigma_{R} = \{ R/\tr(\Omega_{R}^{''-1})\} \Omega_{R}^{''-1}$. Finally, we set the random departure covariance $\Sigma_\Psi = T^{-1}\left(\begin{smallmatrix} 6 & 3\\3 & 9 \end{smallmatrix}\right)$.

\subsection{Estimation results} 
\label{subsec:estimation}

We first evaluate the empirical performance of the parameter estimation. We vary the proportion of the non-zero coefficients $\left\{ [\eta_r]_j \right\}$, i.e., $\omega = {1 - |\Hcal_0|/\{(2p+2)R\}} = \{0.03, 0.05\}$, and set the non-zero entries of $\left\{ [\eta_r]_j \right\}$ equal to 0.5. We repeat the experiment 200 times. To evaluate the estimation accuracy of the covariance matrix $\Sigma^{(r)}$, we report the average and standard error of the bias criterion $\big\{[\hat{\Sigma}^{(r)}]_{b_1,b_2} - [{\Sigma}^{(r)}]_{b_1,b_2} : 1 \le r \le R, |b_1-b_2|\le T \big\}$, as $\Sigma^{(r)}$ and $\hat\Sigma^{(r)}$ are both block-diagonal matrices. To evaluate the estimation accuracy of the regression coefficient $\beta^{(r)}$,  we report the average and standard error of the bias criterion $\big\{ \hat\beta^{(r)}_j - \beta^{(r)}_j: 1 \le r \le R, 1 \le j \le 2p+2 \big\}$. We report the results for $T=4$ in Table \ref{tab:est_T4_2}, and the results for $T=8$ in Table S1 of the supplement in the interest of space. We observe from these tables that, the bias and standard error of the covariance estimation are larger for a larger $\omega$, while those of the regression coefficient estimation do not change much. This matches our theoretical convergence rate in Theorem \ref{lm:rate}, as a larger $\omega$ represents a larger $s_B$ and $c_R$, thus a larger $\theta_{N,T,R,B}$. In addition, as the sample size increases, both the bias and the standard error of the coefficient estimation decrease.

\subsection{Global testing results}

We next evaluate the performance of the global testing procedure. To evaluate the size of the test, we set $\omega = 0$, whereas to evaluate the power of the test, we set $\omega = 0.05$ and set the non-zero entries of $\left\{ [\eta_r]_j \right\}$ equal to 0.2. We set the significance level $\alpha=0.05$. Table \ref{tab:global1} reports the empirical size and power, in percentages, based on 2000 data replications. We see that the empirical size is close to the significance level under all settings, especially for a larger sample size $N$. In contrast, the testing method based on REML has serious size inflation in most settings. For those cases where REML controls the size relatively well, our method achieves a better power. In addition, we also observe that the proposed method achieves a notable power gain when the sample size $N$ or the number of time points $T$ increases, which again agrees with our theoretical findings.

\subsection{Multiple testing results}

We next evaluate the performance of the multiple testing procedure. We adopt the same setting as in Section \ref{subsec:estimation}, and set the pre-specified FDR level $\alpha = 0.1$. We report the empirical FDR and power in Table \ref{tab:multiple_T4_2} based on 200 data replications with $T=4$, and in Table S2 of the supplement with $T=8$. We see from these tables that, for the empirical FDR, the proposed method has FDR well under control across all settings, while the testing method based on REML has some FDR inflation for the cases with a small sample size and a large dimension, e.g., when $N=100$ and $R=100$. For the empirical power, the proposed method is in general more powerful than REML, especially when both methods have the FDR under control. In addition, we also observe that the empirical power of our method increases and the power gain is more apparent when $N$ or $T$ increases.

\section{Longitudinal Neuroimaging Analysis}
\label{sec:realdata}

Alzheimer's disease (AD) is an irreversible neurodegenerative disorder characterized by progressive impairment of cognitive functions, then global incapacity and ultimately death. It is the leading form of dementia, and is currently affecting 5.8 million American adults aged 65 years or older. Its  prevalence continues to grow, and is projected to reach 13.8 million by 2050 \citep{AD2020}. It is thus crucial to better understand, diagnose, and treat this disorder \citep{Jagust2018-review}. We analyze a dataset OASIS-2 from Open Access Series of Imaging Studies (\url{www.oasis-brains.org}). The data consists of a longitudinal collection of 150 subjects aged 60 to 96. Each subject was scanned on two or more visits, separated by at least one year for a total of 373 imaging sessions. For each subject, multiple T1-weighted MRI scans measuring brain gray matter volume were obtained. The subjects are all right-handed and include both men and women. Among them, 72 were characterized as nondemented throughout the study, and 64 were characterized as demented at the time of their initial visits and remained so for subsequent scans \citep{Marcus2010-OASIS2}. We process the data and only include in our data analysis the subjects with $T = 3$ time points and meeting the quality control criteria. This results in $N = 56$ subjects. We then further process the MRI images and parcellate the brain into $R=68$ regions-of-interest (ROIs) using the Desikan-Killiany atlas \citep{desikan2006automated}. For each subject, we also include the binary AD status, sex, education, and socioeconomic status as the time-invariant covariates with $p=4$, and the mini-mental state examination score \citep{folstein1975mini}, atlas scaling factor \citep{buckner2004unified}, and estimated total intracranial volume \citep{buckner2004unified} as the time-variant covariates with $q=3$. 

\begin{figure}[t!]
\centering
\begin{tabular}{ccc}
\includegraphics[width=0.32\textwidth, height=1.8in]{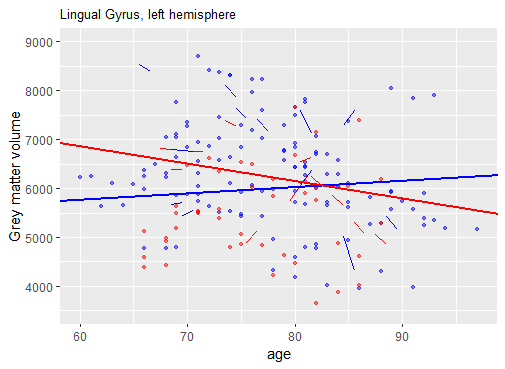} & 
\includegraphics[width=0.32\textwidth, height=1.8in]{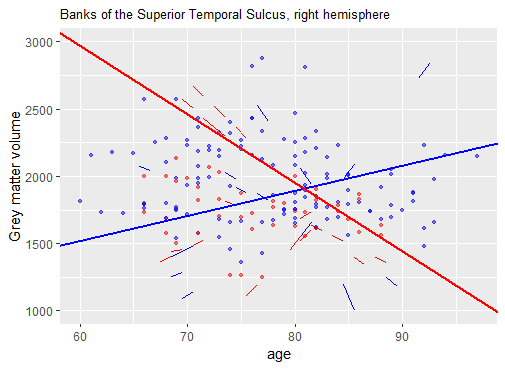} & 
\includegraphics[width=0.32\textwidth, height=1.8in]{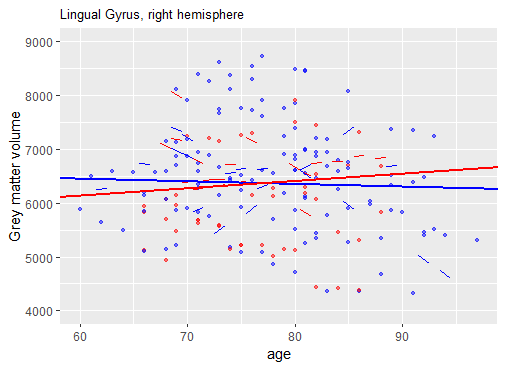}
\end{tabular}
\caption{Scattershot for grey matter volume versus age in three identified brain regions. The short lines denote the individual growth curves of 20 randomly selected subjects from both the AD group (red) and the healthy group (blue). The long bold lines denote the overall growth curves of the two groups.}
\label{fig:real_1} 
\end{figure}

We apply the proposed method to this data, with the pre-specified FDR level $\alpha=0.05$. The test identifies 50 significant coefficients among the total of 680 coefficients. We focus on the ones associated with the binary AD status, and three brain regions are identified, i.e., the lingual gyrus of the left hemisphere, the lingual gyrus of the right hemisphere, and the banks of the superior temporal sulcus of the right hemisphere. Figure \ref{fig:real_1} plots the estimated mean growth curves for individual subjects and the overall trend in those identified regions. We see that, for the lingual gyrus of the left hemisphere and the banks of the superior temporal sulcus of the right hemisphere, both the intercept and slope coefficients differ significantly between the groups of subjects with AD and without, suggesting different starting values as well as different decaying rates. Meanwhile, for the lingual gyrus of the right hemisphere, only the slope coefficient differs significantly between the two groups, suggesting a different decaying rate for the AD patients. These identified brain regions also agree with the AD literature well. In particular, the lingual gyrus is located in the occipital lobe, primarily in the visual processing areas of the brain. Its primary functions include visual processing, visual memory, and visual recognition. It is found associated with emotional processing and visual hallucinations under certain neurological conditions. The superior temporal sulcus is located within the temporal lobe. It plays a significant role in a variety of cognitive and perceptual functions, including processing of auditory and speech information, narrative comprehension, social cognition, among others. \citet{yang2019study} found that the cortical thickness of both lingual gyrus regions demonstrate significance between the AD patients and healthy controls. \citet{guo2020detecting} found that banks of the superior temporal sulcus is the highest beta amyloid affected region, where beta amyloid is one of the most prominent pathological proteins of AD \citep{jack2013tracking}.

\section{Discussion}
\label{sec:discussion}

In this article, we have proposed a new set of estimation and inference procedures for the high-dimensional multi-response GCM. It fills an existing gap in the literature and helps address an important family of scientific questions studying the longitudinal growth trend. Meanwhile, the methodological innovations mainly lie in the proposed multi-step estimation approach and the establishment of the proper convergence rates of the covariance estimators. The former enables us to effectively utilize the data information, while the latter allows us to obtain the desired theoretical guarantees for the proposed tests.

There are several potential extensions of the current proposal. Due to our targeting motivation examples, we consider a large number of response variables, but a relatively small number of predictor variables in our setting. It is possible to extend to high-dimensional predictors as well. Moreover, we focus on a linear type model, which we believe is a good starting point. It is also warranted to consider a nonlinear type GCM. We leave these extensions as future research.

\bibliographystyle{apa}
\bibliography{ref-gcm}


\newpage

\begin{sidewaystable}[t!]
\centering
\caption{Parameter estimation: the bias and standard error based on 200 data replications for the autoregressive and moving average temporal structures with $T=4$.}
\label{tab:est_T4_2} {
\begin{tabular}{c@{\hspace{2em}}c@{\hspace{2em}}c@{\hspace{2em}}c@{\hspace{2em}}c@{\hspace{2em}}c@{\hspace{2em}}c@{\hspace{2em}}c@{\hspace{2em}}c@{\hspace{2em}}c@{\hspace{2em}}c} \toprule
\multicolumn{3}{c}{Temporal structure}         & \multicolumn{4}{c}{Autoregressive}  & \multicolumn{4}{c}{Moving average}\\
	\cmidrule(r){4-7} \cmidrule(r){8-11} 
&         &          & \multicolumn{2}{c}{$\omega=0.03$}  & \multicolumn{2}{c}{$\omega=0.05$} & \multicolumn{2}{c}{$\omega=0.03$}&\multicolumn{2}{c}{$\omega=0.05$}\\  
\cmidrule(r){4-5} \cmidrule(r){6-7}  \cmidrule(r){8-9}  \cmidrule(r){10-11} 
$R$   & $N$ & method   & hub   & small     & hub   & small      & hub   & small          & hub   & small   \\ \midrule
\multicolumn{11}{c}{Bias and SE of covariance estimation}                                                                           \\  \midrule
			\multirow{4}{*}{$50$} &   \multirow{2}{*}{$100$}  & Bias         &0.0798&0.0653&0.1487&0.1411&0.0817&0.0657&0.1467&0.1455
			\\  [2pt] 
			&    & SE       &0.5043&0.3680&0.5206&0.3879&0.4886&0.3727&0.5054&0.3941
				\\  [2pt] 
			&              \multirow{2}{*}{$200$}  & Bias     &0.0742&0.0669&0.1205&0.1151&0.0748&0.0664&0.1205&0.1151
				\\ [2pt] 
			&  & SE    &0.2502&0.2175&0.3255&0.2328&0.2587&0.2291&0.3229&0.2444 \\ [2pt]  \cmidrule(r){1-11}
			
			\multirow{4}{*}{$100$} &         \multirow{2}{*}{$100$}  & Bias           &0.0935&0.0887&0.1540&0.1484&0.0960&0.0909&0.1554&0.1504
				\\ [2pt] 
			&    &SE         &0.4200&0.3787&0.4767&0.4531&0.4211&0.3839&0.4718&0.4520
				\\  [2pt] 
			&        \multirow{2}{*}{$200$}  &Bias           &0.0831&0.0742&0.1342&0.1328&0.0838&0.0747&0.1336&0.1331
				\\  [2pt] 
			&    & SE       &0.3638&0.2244&0.2666&0.2401&0.3538&0.2363&0.2753&0.2520
				\\  [2pt] 
\midrule
\multicolumn{11}{c}{Bias and SE of coefficient estimation}                                                                          \\   \midrule
			\multirow{4}{*}{$50$} &   \multirow{2}{*}{$100$}  & Bias         &  0.0002&-0.0006& 0.0005& 0.0001& 0.0003&-0.0006& 0.0005& 0.0000
				\\  [2pt] 
			&    & SE       &0.1682&0.1688&0.1688&0.1686&0.1672&0.1682&0.1677&0.1675
				\\  [2pt] 
			&              \multirow{2}{*}{$200$}  & Bias     &-0.0002& 0.0002& 0.0001& 0.0003&-0.0002& 0.0002& 0.0001& 0.0003
				\\ [2pt] 
			&  & SE    &0.1103&0.1101&0.1106&0.1106&0.1097&0.1094&0.1099&0.1099
				\\ [2pt]  \cmidrule(r){1-11}
			
			\multirow{4}{*}{$100$} &         \multirow{2}{*}{$100$}  & Bias           & 0.0002& 0.0000& 0.0005&-0.0003& 0.0001& 0.0000& 0.0005&-0.0003
				\\ [2pt] 
			&    &SE         &0.1688&0.1688&0.1689&0.1689&0.1677&0.1678&0.1678&0.1678
				\\  [2pt] 
			&        \multirow{2}{*}{$200$}  &Bias           &0.0000&-0.0001& -0.0001&-0.0001& 0.0000&-0.0001&-0.0001&-0.0001
				\\  [2pt] 
			&    & SE       &0.1108&0.1104&0.1104&0.1106&0.1102&0.1097&0.1097&0.1099
				\\  [2pt] 
			\bottomrule
		\end{tabular}
}	
\end{sidewaystable}

\begin{table}[t!]
\begin{center}
\caption{Global testing: the empirical size and power in percentage based on 2000 replications for the autoregressive and moving average temporal structures with $T = 4$ and $8$, and the significance level $\alpha = 0.05$.}
\label{tab:global1}
\begin{tabular}{c@{\hspace{1em}}c@{\hspace{1em}}c@{\hspace{1em}}c@{\hspace{1em}}c@{\hspace{1em}}c@{\hspace{1em}}c@{\hspace{1em}}c@{\hspace{1em}}c@{\hspace{1em}}c@{\hspace{1em}}c}
				\toprule
				\multicolumn{3}{c@{\hspace{1em}}}{}      &    \multicolumn{4}{c@{\hspace{1em}}}{$T=4$} & \multicolumn{4}{c}{$T=8$}  \\ 
				\multicolumn{3}{c@{\hspace{1em}}}{}      &    \multicolumn{2}{c@{\hspace{1em}}}{Empirical Size} & \multicolumn{2}{c@{\hspace{1em}}}{Empirical Power} &    \multicolumn{2}{c@{\hspace{1em}}}{Empirical Size} & \multicolumn{2}{c}{Empirical Power}  \\  \cmidrule(r){4-5} \cmidrule(r){6-7}  \cmidrule(r){8-9} \cmidrule(r){10-11}
				\multicolumn{1}{c@{\hspace{1em}}}{$R$}   & \multicolumn{1}{c@{\hspace{1em}}}{$N$} & \multicolumn{1}{c@{\hspace{1em}}}{Method}   & \multicolumn{1}{c@{\hspace{1em}}}{hub}   & \multicolumn{1}{c@{\hspace{1em}}}{small}    & \multicolumn{1}{c@{\hspace{1em}}}{hub}  & \multicolumn{1}{c@{\hspace{1em}}}{small}    & \multicolumn{1}{c@{\hspace{1em}}}{hub}   & \multicolumn{1}{c}{small}  & \multicolumn{1}{c@{\hspace{1em}}}{hub}   & \multicolumn{1}{c}{small}     \\ \midrule
				\multicolumn{11}{c}{Auto-regressive temporal structure} \\  \midrule
				
				\multirow{4}{*}{$50$} &   \multirow{2}{*}{$100$}      &Proposed    &5.6&4.2&20.5&15.7&5.4&6.0&52.4&58.1 \\ [2pt] 
				& & REML& 9.5&9.8&25.1&23.6& 11.1&9.0&49.6&53.0 \\ [2pt] 
				&   \multirow{2}{*}{$200$}      &Proposed      & 4.3&5.0&58&55.1& 5.0 &5.9&99.4&99.9\\ [2pt] 
				&&REML &6.0&6.2&56.1&55.5 & 6.9&7.0&98.6&98.3\\ [2pt] \cmidrule(r){1-11}
				
				\multirow{4}{*}{$100$} &   \multirow{2}{*}{$100$}     &Proposed       & 4.3&4.4&17.7&17.3 &5.8&5.0&74.5&65.3\\[2pt]  
				&&REML&  9.8&11.1&27.4&29&  12.2&10.6&69.2&67.6 \\ [2pt] 
				&   \multirow{2}{*}{$200$}     &Proposed       &4.3&3.9&61.1&67.1  &5.1&4.8&100.0 &99.9\\ [2pt] 
				&&REML& 7.0&7.6&65.4&68.3 &  7.2&7.0&99.7&99.4\\ [2pt] 
				\midrule
				
				\multicolumn{11}{c}{Moving average temporal structure}      \\ \midrule
				\multirow{4}{*}{$50$} &   \multirow{2}{*}{$100$}      &Proposed    &  5.6&4.2&21.2&14.6  & 6.2&7.2&54.9&64.1\\ [2pt] 
				& & REML& 9.7&9.2&24.6&23.4 & 10.8&10.0&47.8&51.7\\ [2pt] 
				&   \multirow{2}{*}{$200$}      &Proposed      &4.7&4.7&58.3&54.9   &5.1&6.1&99.5&99.8 \\ [2pt] 
				&&REML & 6.1&5.9&54.8&54.2& 6.9&7.6&97.9&97.4 \\ [2pt]  \cmidrule(r){1-11}
				
				\multirow{4}{*}{$100$} &   \multirow{2}{*}{$100$}     &Proposed       & 4.7&4.0&17.7&16.4 & 6.7&5.7&79.0&68.3\\[2pt]  
				&&REML&  10.0&11.1&27.4&28.0 &  12.2&11.1&66.6&65.1 \\ [2pt] 
				&   \multirow{2}{*}{$200$}     &Proposed       &  4.2&3.7&61.1&67.2 & 5.3&5.2&100.0 &100.0\\ [2pt] 
				&&REML&7.0&7.9&63.7&66.4&7.6&7.1&99.4&99.0\\
				\bottomrule
			\end{tabular}
\end{center}
\end{table}

\begin{sidewaystable}[t!]
\centering
\caption{Multiple testing: the empirical FDR and power in percentage based on 200 data replication for the autoregressive and moving average temporal structures  with $T=4$ and the FDR level $\alpha=0.1$.}
\label{tab:multiple_T4_2}{
\begin{tabular}{c@{\hspace{2em}}c@{\hspace{2em}}c@{\hspace{2em}}c@{\hspace{2em}}c@{\hspace{2em}}c@{\hspace{2em}}c@{\hspace{2em}}c@{\hspace{2em}}c@{\hspace{2em}}c@{\hspace{2em}}c} \toprule
\multicolumn{3}{c}{Temporal structure}         & \multicolumn{4}{c}{Autoregressive}  & \multicolumn{4}{c}{Moving average}\\
\cmidrule(r){4-7} \cmidrule(r){8-11} 
&         &          & \multicolumn{2}{c}{$\omega=0.03$}  & \multicolumn{2}{c}{$\omega=0.05$} & \multicolumn{2}{c}{$\omega=0.03$}&\multicolumn{2}{c}{$\omega=0.05$}\\  
\cmidrule(r){4-5} \cmidrule(r){6-7}  \cmidrule(r){8-9}  \cmidrule(r){10-11}
$R$   & $N$ & method   & hub   & small     & hub   & small      & hub   & small          & hub   & small   \\ \midrule
\multicolumn{11}{c}{Empirical FDR}                                                                           \\  \midrule
			\multirow{4}{*}{$50$} &   \multirow{2}{*}{$100$}  & Proposed         &6.82&9.23&7.47&7.07&7.06&7.02&7.18&7.01 \\  [2pt] 
			&    & REML        &9.46& 8.85&11.83&12.34& 9.35& 8.27&11.76&12.58 \\  [2pt] 
			&              \multirow{2}{*}{$200$}  & Proposed      &7.65&7.69&7.52&7.44&7.86&7.66&7.33&7.47 \\ [2pt] 
			&  & REML     &10.42&10.67&10.99&11.18&10.39&10.51&10.78&11.23  \\ [2pt]  \cmidrule(r){1-11}
			
			\multirow{4}{*}{$100$} &         \multirow{2}{*}{$100$}  & Proposed           & 7.05&7.85&6.78&6.85&7.02&7.60&6.68&6.73 \\ [2pt] 
			&    & REML         &12.48&13.18&11.74&11.96&12.53&13.31&11.78&11.89 \\  [2pt] 
			&        \multirow{2}{*}{$200$}  & Proposed           & 7.17&7.99&6.85&6.56&7.33&8.07&6.79&6.68 \\  [2pt] 
			&    & REML        &10.57&10.94&10.58&10.25&10.39&10.89&10.59&10.22 \\  [2pt] 
			\midrule			
\multicolumn{11}{c}{Empirical Power}                                                                            \\   \midrule
			\multirow{4}{*}{$50$} &   \multirow{2}{*}{$100$}  & Proposed    &34.98&35.98&48.43&47.04&37.09&35.68&48.71&47.48 \\  [2pt] 
			&    & REML        &36.55&34.42&49.78&48.84 &35.89&33.32&49.54&48.55 \\  [2pt] 
			&              \multirow{2}{*}{$200$}  & Proposed     &91.52&91.67&90.68&92.54 &92.21&92.15&91.41&93.01 \\ [2pt] 
			&    & REML      &89.14&88.80&89.20&90.84&89.82&89.42&89.85&91.29 \\  [2pt]  \cmidrule(r){1-11}
			
			\multirow{4}{*}{$100$} &         \multirow{2}{*}{$100$}  & Proposed          & 39.76&40.09&47.05&46.87&40.29&40.54&47.49&47.36 \\ [2pt] 
			&    & REML       &41.23&41.44&49.55&49.33&41.61&41.38&49.61&49.59\\  [2pt] 
			&        \multirow{2}{*}{$200$}  & Proposed           &90.45&91.20&92.53&93.10&90.95&91.74&93.05&93.59 \\ [2pt] 
			&    & REML       &88.70&88.34&91.32&91.60&89.31&89.15&91.89&92.06 \\ 
			\bottomrule
\end{tabular}
}	
\end{sidewaystable}

\end{document}